\documentclass[%
 prb, jmp, amsmath,amssymb,
reprint,%
]{revtex4-1}

\usepackage{graphicx}
\usepackage{booktabs}
\usepackage{dcolumn}% Align table columns on decimal point
\usepackage{bm}

\begin{document}
\preprint{}

%opening
\title{Fluid-solid transition in hard hyper-sphere systems.}

%AIP
\author{C.D. Estrada}\email[]{cdea@cie.unam.mx}
\author{M. Robles}\email[]{mrp@cie.unam.mx}
%\thanks{}
\affiliation{Centro de Investigaci\'on en Energ\'ia, UNAM\\Priv. Xochicalco S/N, Col. Centro\\62580, Temixco Mor. M\'exico.}
\date{\today}


%\begin{document}\maketitle
\begin{abstract}
In this work we present a numerical study, based on molecular dynamics simulations, to estimate
the freezing point of hard spheres and hypersphere systems in dimension $D=4, 5, 6$ and $7$.
We have studied the changes of the Radial Distribution Function (RDF) as
a function of density in the coexistence region. We started our simulations from
crystalline states with densities above the melting point, and moved down to
densities in the liquid state below the freezing point. 
For all the examined dimensions (including $D=3$) it was observed
that the height of the first minimum of the RDF changes in an almost continuous way
around the freezing density and resembles a second order phase transition. 
With these results we propose a numerical method to estimate the freezing point as a function of the 
dimension $D$ using numerical fits and semiempirical approaches. We find that the 
estimated values of the freezing point are very close to
previously reported values from simulations and theoretical approaches 
up to $D=6$ reinforcing the validity of the proposed method. 
This was also applied to 
numerical simulations for $D=7$ giving new estimations of the freezing point for this dimensionality.
\end{abstract}

\maketitle

\section{Introduction}

Although the study of fluids of hard spheres in dimensions higher than three is not new \cite{RH64, Lthss, MT, Frsr, LubanB, Jsln} 
it has attracted renewed attention in recent years \cite{AM08}.
This interest is because as we change the dimension these systems share common properties 
that may lead to general relations among them and eventually could bring new insights to traditionally
difficult problems for two and three dimensional systems. 

One of such properties is the fluid-solid transition, which has been widely studied in hard-sphere fluids and is known to appear also for hard-hypersphere fluids. Evidence of this transition 
found by numerical simulations\cite{MT} and theoretical estimations\cite{FNK}  
suggests that it preserves the same characteristics of the three dimensional case. Nevertheless, the exact
determination of freezing and melting points at
any dimension (including two and three) is, at present, not possible. Only using numerical simulations 
and theoretical approaches is how these values are known with some accuracy. 

The freezing point of the hard-sphere (HS) fluid was first observed and estimated by Alder and Wainright in 1957
using molecular dynamic simulations \cite{AW,AW2,AW3}. Since then the problem has been revised using
different simulation techniques, ranging from Monte Carlo methods \cite{WJ,RH,NVM} to the recent Direct Coexistence Simulations \cite{NVM}. 
%Also different theoretical approaches based on Density Functional Theories (DFT) \cite{PT85,WAC85, DA,LB} and Mean-Field Cage Theories (MFCT)  \cite{XZW}, have been applied to the same problem. 
Also different theoretical approaches have been applied to the same problem, such as Mean-Field Cage Theories (MFCT)  \cite{XZW} and Density Functional Theories (DFT) \cite{PT85,WAC85, DA,LB}, in particular the fundamental measure theory (FMT) by Rosenfeld \cite{YR89} describes both the fluid and solid phases while keeping the correct limit in the close packing. Another advantage is that, in contrast to other density functional approaches, it does not require the direct correlation function as an input.
All of the above works seem to coincide that freezing occurs at a packing fraction $\eta$ close to $\eta_F\approx0.494$. This observation has been also confirmed by different experiments with colloidal suspensions of hard-core particles\cite{CCRMZRO, BWQSPP}. 

For hard-hypersphere(HHS) fluids at dimensions $D>3$, the information available is so far
limited. In 1984, Leutheusser \cite{Lthss} found an analytic solution, using the Percus-Yevick approximation for the HHS problem that allows one  to obtain algebraic expressions for the pressure equation of state in odd dimensions. At present such equation of state has been computed only up to $D=7$ using the compressibility and the virial routes\cite{R-LH2}. Unfortunately, the Percus-Yevick approximation gives no information on the freezing
transition and neither do other empirical and semiempirical approximations to the equation
of state. Recently, some theoretical approaches have been developed 
for the prediction of the freezing transition as a function of $D$. For example, Finken et al. \cite{FNK} 
generalized the scaled-particle theory to arbitrary $D$ for the fluid phase and a cell theory for the crystalline phase to predict a first order freezing transition up to $D=50$. In the
same direction Wang \cite{XZW}, based on a MFCT, has obtained a simple expression for the
freezing point as a function of $D$. At the moment, a full validation of both predictions based on
computational simulations can not be done because of the lack of results. 
As far as we know, numerical simulations have been carried out to obtain information on the fluid equation of state for $D=4$ to $D=9$ \cite{GGM91,MT,R-LH2,Bsh2,BCW}, but the freezing points have only been estimated for  
$D=4,5$ \cite{MT, Skg, VMeel} and recently Van Meel et al. \cite{ VMeel2} provided more accurate estimations up to $D=6$, while for $D=7$ \cite{R-LH2}, the phase transition has been reported but its location has not been estimated.

The main purpose of this work is to propose a simple empirical method, based on molecular dynamics simulations, 
to estimate quantitatively the freezing and melting point for HHS fluids.
Our proposal focuses on the changes in the Radial Distribution Function (RDF) when a
fluid-solid transition takes place. For $D=2$ and $D=3$ it is well established that the RDF must contains relevant
information on the structural precursors of freezing \cite{TRSK} and therefore an empirical parameter
defined on the basis of the changes in shape could be used to find the freezing point
with some accuracy. 
Some empirical rules have been previously proposed for three dimensional fluids, like the Hansen-Verlet crystallization
rule \cite{JPH69}, based on the height of the fist peak of the static structure factor. Also
 in 1978 Wendt and Abraham \cite{Abrh} used the changes in the structure of the second
peak in the RDF for metastable states to define a parameter that could allow them to estimate the glass 
transition density. 
Actually the properties of the second peak of the RDF as density increases are directly related with the crystallization processes and the appearance of a shoulder on it is considered as a signature of the fluid solid transition. The study of such changes for fluids at higher dimensions may give some criteria to estimate the freezing point on them, nevertheless as dimensionality increases the oscillation of $g(r)$ gets significantly damped \cite{BshpStr,Skg}, and hence the characteristic shoulder of the second peak becomes difficult to appreciate.
Based on this idea, we decided to use the properties of the first minimum of the RDF instead, to
find the freezing point for $D>3$.

The work is organized as follows: in section \ref{sec2} a review of the fluid-solid transition is
included, as well as new molecular dynamics numerical simulations for the HS system at packing fractions 
close to and between the freezing and the melting points. This allows us to 
define and validate the method we will later generalize to find the freezing and melting points of HHS
systems for $D>3$. In section \ref{sec3} we report the results obtained with that numerical proposal for dimensions
$D=4,..,7$. A discussion of the obtained results is included in section \ref{sec4} and the paper is closed with a summary of the results and further conclusions in section \ref{sec5}.

\section{Fluid-Solid transition in the Hard Sphere System.}
\label{sec2}

The HS system is defined as a set of $N$ identical molecules interacting with the hard-core potential:
\begin{equation}
U_{HS}(r_{ij})=\left\{ \begin{array}{c c r}
\infty & & r_{ij}<\sigma \\
0 & & r_{ij}\geq \sigma \\
        \end{array}
\right.
\end{equation}
where $\sigma$ is the diameter of the spheres.\\

Is well known that, because of the form of the interaction, the compressibility factor of the HS fluid: $Z=PV/N K_B T$ (where $P$ is the pressure, $N$ the number of particles, $V$ the system volume, $T$ the temperature and $K_B$ the Boltzmann constant) can be written as a function of only one variable, either the reduced density $\rho=N \sigma^3/V$ or the packing fraction $\eta=\rho V_{sph}$, with $V_{sph}$ the volume of a unit diameter sphere. For a three dimensional HS fluid
the volume of the unitary sphere is simply given by $V_{sph}=\pi /6$.

The phase diagram can be reduced to only two stable phases: fluid and crystal, defined completely by the freezing and melting packing fractions, $\eta_F$ and $\eta_M$ respectively. With
some accuracy is possible to say that these packing fractions are located at $\eta_F=0.494$ and $\eta_M=0.545$
(see Table \ref{TFMP3D}) respectively.

The pressure equation of state as a function of $\eta$ draws a stable 
fluid branch and extends up to the freezing point, where the freezing pressure remains constant
up to the melting point. 
At this point, the crystal branch appears 
extending up to the close packing density ($\eta_{CP}=\pi/3\sqrt{2}$). 
Compressing the hard sphere fluid beyond the freezing point but preventing crystallization, 
the system may enter into a metastable fluid branch.
It has also been widely studied and it is known to drive the fluid phase to a supercooled liquid like states 
and undergo a glass transition.

The freezing transition implies qualitative and quantitative changes in the structural properties,
that may be observed in the Radial Distribution Function (RDF). 
One characteristic effect observed in the HS fluid,
is the appearance of a shoulder in the second peak for dense fluid states close to the freezing point,
which is associated with the formation of local crystalline regions \cite{TRSK}. Another, perhaps less
examined feature, is a change in the width and depth of the first minimum, that could be related with 
the cage effect produced by the local crystalline order.

Our proposal here, is to examine using Molecular Dynamics simulations the changes in the first minimum
of the RDF with the purpose of 
defining a method to measure, with reasonable accuracy, the freezing point
in the three dimensional HS fluid. We believe that numerically this procedure may be simpler and more accurate
than a method based on the evolution of the shoulder of the second peak. In addition this could also be 
simple to extended to hard hyper-sphere fluids at arbitrary dimensions. 

\subsection{Simulation Details.}

In order to design a general simulation procedure common to HS and HHS, we decided to start examining the three
dimensional system, simulating states in the crystal branch from dense crystal to dense fluid phase. The positions of spheres were set initially in a face centered cubic (FCC) structure with
periodic boundary conditions and initial velocities chosen from a Maxwell Boltzmann distribution. To establish a balance among simulation time and reliability
of averages, we set the initial positions in a FCC crystal of $6\times6\times6$ unitary cells, giving
a total of $846$ particles. Since the number of particles and simulation volume cell are fixed, 
the diameter was used to change the packing fraction and therefore the density.
In addition, in this particular case, for comparison purposes we have used previous simulations for the metastable fluid branch which use $3951$ particles in dense random packed states \cite{Tesis}.

Based on the algorithm of Alder and Wainwright\cite{AW2} to simulate the Hard Spheres fluid  we
performed event-driven molecular dynamics simulations, moving the system in successive collisions. 
The equation of state
was measured computing the compressibility factor through the mean virial forces, given by the change on momentum  at collision, as:
\begin{equation}\label{Zdef}
\Xi=-\frac{1}{N\langle v^2\rangle t }\sum_{c=1}^{N_c(t)}\vec{r}_{i,j}\cdot \Delta \vec{v}_i,
\end{equation}
which is related to the compressibility factor through $Z=1+\Xi$. In Eq. (\ref{Zdef}) $N$ is the number of particles, $\langle v^2\rangle$ is the mean square velocity, $t$ is the time, $N_c(t)$ is the total number of collisions at $t$, $\vec{r}_{i,j}$ is the vector joining the center of the colliding particles ${i,j}$ and $\Delta \vec{v}_i$ is the change in velocity of particle $i$.
For the structural properties, RDF was obtained through the standard procedure by determining the mean number of particles at a distance between $r$ and $r+dr$ away from one of them, averaging over all particles and events.

In order to obtain equilibrium properties, before the final simulation a stabilization stage was performed, where the system is left to reach the equilibrium ageing by one hundred thousand collisions and verifying that the change in the mean virial forces 
reaches a stable mean value.
Once the system is stable the final configuration was taken as the initial condition for the true simulation. 
The final simulation stage was proposed to take averages of $100,000$ collision steps.

\subsection{Equation of state, structural properties and the phase transition.} 

\begin{figure}
\includegraphics[scale=1]{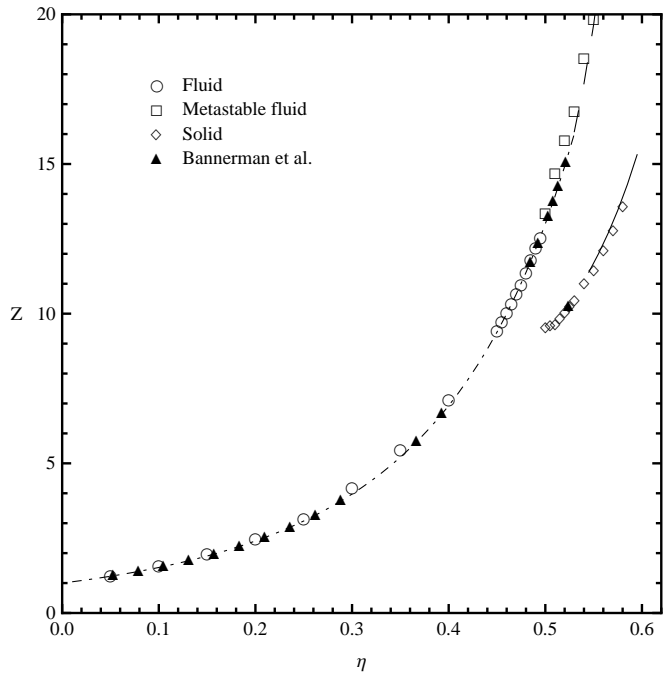}
\caption{Compressibility factor $Z$ for the Hard Sphere system.
The chosen simulation states for each phase are denoted by circles (fluid), squares (metastable states) 
and diamonds (crystal), simulation results by Bannerman ($\blacktriangle$) and the lines denote the equations of state: Carnahan-Starling (dot-dashed) \cite{CS} , Speedy (dashed) \cite{Spd} 
and Hall (continuous) \cite{Hll}.}\label{Z3D}
\end{figure}

The simulation results for the three dimensional case were validated using some analytic approaches
for each branch. For the states in the fluid and metastable fluid phases we compared with the Carnahan-Starling 
equation of state \cite{CS}. The crystal phase was approached with the Hall equation \cite{Hll} and
for the glassy state we used the free volume approach from Speedy \cite{Spd}.
As can be seen in Fig. \ref{Z3D}, numerical results are in good agreement with analytic equations, giving enough evidence 
that the simulation program is working properly.  
Although the used equations correspond to simple empirical expressions, they provide a good 
approximation to the general behavior of the system. Recent efforts by Bannerman et al.\cite{MNB10} 
have been made in order to obtain a new and more accurate equation of state, also compatible with 
the known high order virial coefficients.% A discussion of these proposals were made by Bannerman et al. \cite{MNB10}. 
This reference also provides recent accurate compressibility data from simulations with large 
systems of about $10^5$ and $10^6$ particles, which are compatible with our simulation data (see Fig. \ref{Z3D}).

It is worth to mention that, from the same figure, it is possible to note that molecular dynamics drive the 
system from stable crystal to metastable crystal states for densities below the melting point. Such states
are reproducible under the same simulation conditions, and the error bars which lie within 1\% were not included to avoid overcrowding.

Also for validation purposes
the RDF was measured directly from simulations and compared with predictions from the Rational Function
Approximation (RFA) method  proposed by Yuste and Santos\cite{RFA}. 
To compare with our simulations we used the Carnahan-Starling equation of state as input in the RFA scheme, predicting with 
excellent agreement the RDF's in the fluid phase but loosing accuracy close to the freezing point.

\begin{figure*}
\includegraphics[scale=1]{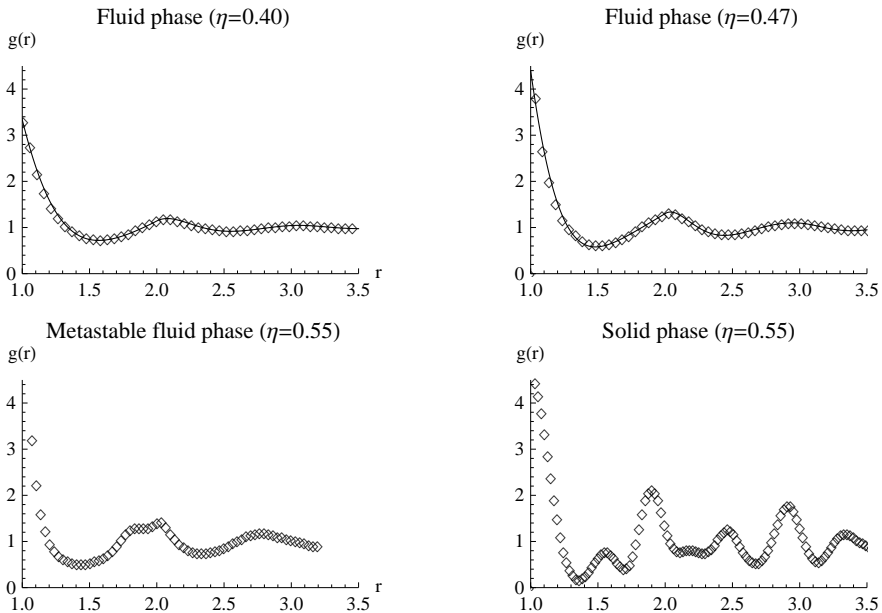}
\caption{Simulation (points) and theoretical (line) Radial distribution Function for densities corresponding to fluid, solid and metastable states.}\label{FGR3D}
\end{figure*}

Figure \ref{FGR3D} shows some results for the RDF in the fluid, metastable and solid phases. For states just below the freezing point ($\eta=0.47$), the simulated RDF shows the characteristic shoulder  
on the second peak, indicating the rising of preceding structures of crystallization. Finally in the crystalline phase the RDF shows the rise of the peaks 
corresponding to the 
FCC structure, in such a way the simulations are giving 
expected results which can be used for more detailed analysis.

\subsection{The RDF analysis.}

The shape of the second peak of the RDF close to the freezing transition has been used in the
prediction or analysis of fluid-solid transitions\cite{TRSK,Abrh}. 
Nevertheless, the changes of the first minimum in the freezing transition, 
as far as we know, has not been deeply examined and can also carry important information.

To measure the position of the first minimum from our simulation results, which we defined located at position $r=X_{min}$ 
and with depth $g(X_{min})$, we decided to fit the first minimum of the RDF with a polynomial function $f(x)$ of the form:
\begin{equation}
 f(x)=\frac{a_0}{x^2}+\frac{a_1}{x}+a_2+a_3 x+ a_4 x^2 +a_5 x^3,
\end{equation}
to determine the coefficients $a_i$ by least squares and to obtain the minimum by
solving the equation $f'(X_{min})=0$.
The mean values of $X_{min}$ and $g(X_{min})$ 
were associated with the values obtained from for the final averaged RDF.
The error bars were obtained computing the mean square difference between the instantaneous  
(at a time step, averaging only over particles) and the mean values in a sample of the last 
1000 configurational states.

Recently, using the short and long 
range asymptotic behavior from the Percus-Yevick approximation, Trokhymchuk et al. \cite{AT05}
 derived an accurate analytical equation for $g(r)$, and the parametrized relations 
\begin{equation}\label{trok-Xmin}
X_{min}=2.0116-1.0647\eta+0.0538 \eta^2
\end{equation}
and
\begin{eqnarray} \label{trok-gXmin}
g(X_{min})&=&1.0286-0.6095\eta +3.5781\eta^2 \\
&& -21.3651\eta^3 +42.6344\eta^4 -33.8485\eta^5 \nonumber
\end{eqnarray}
for the location of the minimum in the range $0.1<\eta<0.47$. These expressions were derived only for $D=3$.

\begin{figure}
\includegraphics[scale=1]{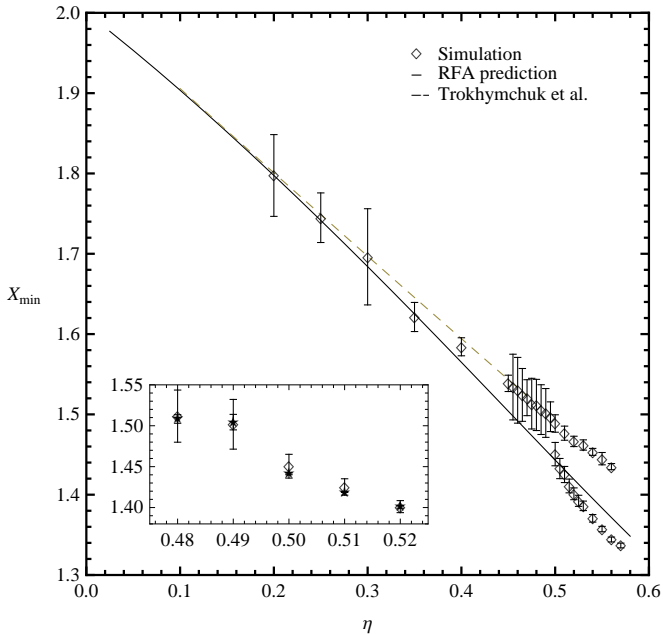}
\caption{Comparison for the position of first minimum $X_{min}$ of RDF for the hard sphere systems. The continuous line is the result from RFA method and the dashed line is the Trokhymchuk et al. expression given in equation (\ref{trok-Xmin}). Diamonds ($\lozenge$) represent our simulation results with 864 particles and in the inset stars ($\star$) correspond to 2048 particles.}\label{Xm3D}
\end{figure}

\begin{figure}
\includegraphics[scale=1]{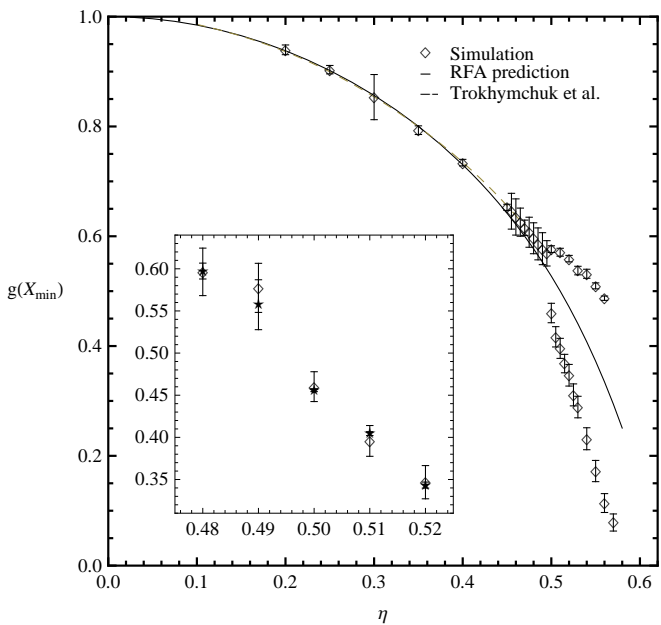}
\caption{Behavior of the value $g(X_{min})$ with packing fraction for hard spheres: From the RFA method (continuous line), Trokhymchuk et al. expression given in equation (\ref{trok-gXmin}) (dashed line) and simulation data for 864 ($\lozenge$) and 2048 particles ($\star$).}\label{GXm3D}
\end{figure}

The simulation results for $X_{min}$ and $g(X_{min})$ are presented as functions of the packing fraction in 
Fig. \ref{Xm3D} and  Fig. \ref{GXm3D}. For comparison, we have included the predictions derived from the
RFA method and Trokhymchuk's expressions represented with continuous and segmented lines in both figures.
Although the RFA predictions for $X_{min}$ are reasonably accurate while expression (\ref{trok-Xmin}) shows a better 
agreement, the result for $g(X_{min})$, is remarkable because the simulation data coincides with both the RFA theory 
and expression (\ref{trok-gXmin}) for the whole fluid branch from low densities to the freezing point. 
Beyond the freezing point, the simulation data split into two, showing a clear difference between the metastable 
fluid and metastable crystal branches resembling a second order phase transition.
It is important to remark that this change in $g(X_{min})$ may not be strictly interpreted as signaling a second order phase transition in a thermodynamic sense i.e: equality of pressure and chemical 
potential in both phases. However, since for hard core systems the phase transition 
is purely entropic, one would expect that the changes in structure such as the one present in $g(X_{min})$ may indeed have a bearing on it. Therefore we assume that
%This result suggests that, 
the changes on the mean value of $X_{min}$ and in particular $g(X_{min})$ are good indicators for the freezing 
transition and that the freezing point can be estimated using interpolation techniques on any of them. However, 
the use of $g(X_{min})$ looks better for that purpose, because the data show a cleaner aspect.\\

We also studied the influence of the system's size by the inspection of the changes on measurements of 
$X_{min}$ and $g(X_{min})$ for a 2048 particles with respect to the 864 particles system. Since the results
obtained with 864 particles seems to reproduce well the theoretical estimations in the fluid branch, we choose to increase the
system size for selected densities in the metastable crystalline branch, which is as well the most important for our
purposes. The comparison is shown in the insets of Fig. \ref{Xm3D} and Fig. \ref{GXm3D}, there we observed a maximal deviation of 2\% on the mean value and a decrease in the error bars as we increase the number of particles more than double. Then we conclude that the shape obtained for both parameters will remain almost the same with an important raise in particle number.

To estimate the freezing point from the simulation data, as a first approximation, we fit the simulation
data of $g(X_{min})$ with two second order polynomials for the fluid and the metastable solid branches. The 
intersection between these polynomials gives an estimation of the freezing point ($\eta_F)$. From the equations 
of state for fluid states, the corresponding coexistence reduced pressure is computed as 
$p_F^*=\eta_F Z(\eta_F) /V_{sph}$ and the extrapolation to the crystalline states determines 
the melting density $\eta_M$. These results were compared with previously reported values in Table \ref{TFMP3D}, 
showing that although the estimation is rough the technique gives comparable results.

\begin{table}
\caption{Estimated values of the freezing $\eta_F$ and melting $\eta_M$ packing fractions as well as the coexistence pressure for the HS system.}\label{TFMP3D}
%\begin{center}
\begin{ruledtabular}
\begin{tabular}{llll} 
  & $\eta_F$ & $\eta_M$ & $p_F^*$ \\ \hline 
MD  \cite{AW3}		&0.494	&0.545	&\\ 
MC  \cite{RH} 		&0.494	&0.545	&11.7 \\ 
    NpT, NpNAT\cite{NVM}	&0.487	&0.545	& 11.54\\
    VSNVT \cite{NVM}	&0.491	&0.544	& 11.54\\ 
MWDA \cite{DA}		&0.476	&0.542	&10.1 \\ 
GELA \cite{LB}		&0.495	&0.545	&11.9 \\ 
EXP.\cite{BWQSPP}	&0.494	&0.545	&\\ 
This Work           & 0.488& 0.545& 10.95\\ 
\end{tabular}
\end{ruledtabular}
%\end{center}
\end{table}

From this analysis we conclude that: 
\begin{enumerate}
 \item Our simulations reproduce the stable branches of the phase diagram, as well as the metastable branches.
 \item The RFA predictions for the position of the first minimum give, compared with simulations, good
qualitative and quantitative results for the fluid phase up to the freezing point. Beyond this point, it is not 
compatible with metastable branches.
 \item The analysis of the behavior of the first minimum on the RDF and the reasonable agreement of the estimated
 values  with previous known results, suggest that these properties can be used to estimate
 the freezing point in an empirical way.    
\end{enumerate}

In the next section we will discuss an extension of the numerical method, to approximate the freezing
point for systems at higher dimensions, examining the first minimum in the RDF of HHS from 4 to 7 dimensions, focusing on a quantitative analysis for the 
value $g(X_{min})$.

\section{The extension to the case of $D$ dimensional hard hyper-sphere.}
\label{sec3}

Based on the computational algorithm to simulate the hard sphere system \cite{AW2}, we made an extension to  $D$-dimensional hard hyper-sphere systems following the same simulation technique used by Michels and Trappeniers \cite{MT} for $D=4$ and $D=5$. The computational code was designed to be extended to any dimension just by changing a library which includes all mathematical operations that depend on dimension. 
In this way the program use exactly the same dynamics on multidimensional spaces.\\

For an arbitrary dimension $D$ the 
generalization of the
%***
reduced density of a collection of hyper-spheres of diameter $\sigma$ 
is defined as $\rho=N \sigma^D/V$ and allows us
to define a generalized packing fraction by $\eta=\rho V_{sphD}$, where  
$V_{sphD}$ is the volume of a hypersphere of unit diameter, which is given by:
\[
V_{sphD}(\sigma=1)=\frac{(\pi/4)^{D/2}}{\Gamma(1+D/2)} \sigma^D
\]
where $\Gamma$ denote the usual Gamma function.

The simulation box used for all numerical experiments is an hypercube of
side $L$ with periodic boundary conditions and volume $V=L^D$. To start
from a crystalline state we choose to divide the simulation hypervolume
in $N_c$ $D-$type primitive hypercells of side $l$ in each direction,
such that $V=(N_c l)^D$. For a $D-$type lattice, the primitive hypercell
contains $N_D=2^{D-1}$ spheres and therefore the number of particles in
the simulation box as a function of $N_c$ and $D$ is $N_p=2^{D-1} N_c^D$.
In Table \ref{tabnc} we present the explicit numbers of particles in the
simulation box for dimensionality between three an seven and $N_c=1$ to $N_c=4$.
Under this scheme the growth in the number of particles is geometrical
and do the simulation very costly as the primitive cells and dimensionality increase.

\begin{table}\caption{Number of particles in a $D$-type lattice as function of the dimensionality $D$ and the hypercubic cell size $N_c$.}\label{tabnc}
\begin{ruledtabular}
\begin{tabular}{lllll}
\multicolumn{1}{c}{$D$}&\multicolumn{4}{c}{$N_c$}\\ 
	      & 1 & 2  & 3   & 4 \\ \hline
3         & 4 & 32 & 108 & 256 \\
4         & 8 & 128 & 648 & 2048 \\
5         & 16 & 512 & 3888 & 16384 \\
6         & 32 & 2048 & 23328 & 131072 \\
7         & 64 & 8192 & 139968 & 1048576
\end{tabular}
\end{ruledtabular}
\end{table}
								
For the present simulations, and again to keep a reasonable balance between computing time and being able to
examine a wide density range, the initial configurations were chosen as shown in Table \ref{tablecell}.

\begin{table}
\caption{Number of hypercells for simulation}\label{tablecell}
\begin{ruledtabular}
\begin{tabular}{cccc} 
 D & $N_c$ & $N_c^D$ & Num. Particles \\ \hline
4 & 2 &16 & 128 \\ 
5 & 2 &32 & 512 \\
6 & 1 &1 & 32 \\
7 & 1 &1 & 64 \\
\end{tabular}             
\end{ruledtabular}
\end{table}

\begin{figure}
\includegraphics[scale=0.8]{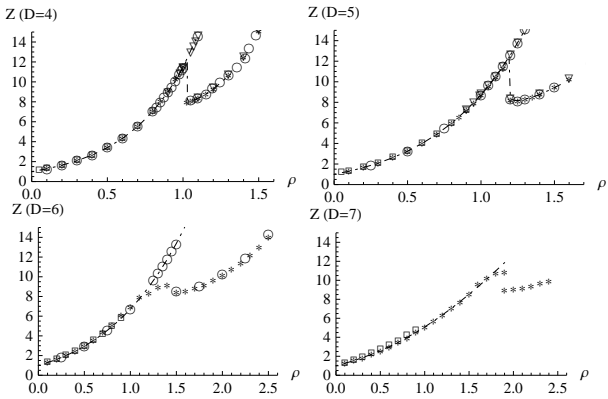}
\caption{Comparison of measured $Z$ $(\star)$ for set of particles given in Table \ref{tablecell} with data from Van Meel et al. ($\circ$), Skoge et al. (continuous line), Bishop et al. ($\square$), Lue et al. ($\triangledown$) and Pad\'es $Z_{[4,5]}$ (segmented line).}\label{BComp}
\end{figure}

To validate our numerical results we considered recent 
simulation data from Skoge et al. \cite{Skg}, van Meel et al. \cite{VMeel2} and Lue et al. \cite{LLMB10}.
The simulations of Skoge et al. were done for 4 and 5 dimensions with a maximum number of particles of 
32768 and 124416, respectively. On the other hand the simulations of van Meel et al. were for 4, 5 and 6 dimensions, using a maximum number of particles of 4096 , 3888 and 2048 for the same $D$-lattices as initial states. 
Finally Lue et al. examined the  solid lattices ($D4$ and $D5$)  from event-driven molecular dynamics (10000 and 16807 particles, respectively), and the metastable fluid phases by Monte Carlo simulations (from 4096 to 10000 and 3125 to 7776 particles, respectively).
%To validate our numerical results we considered recent simulation data from Skoge et al. \cite{Skg} and van Meel et al. \cite{VMeel2} valid between the freezing and the melting points. The simulations of Skoge et al. were done for 4 and 5 dimensions with a maximum number of particles of 32768 and 124416 respectively. In the other hand the simulations of van Meel were for 4, 5 and 6 dimensions, using maximum number of particles of 4096 , 3888 and 2048 particles for the same D-lattices as initial states. 
Compared with these simulations the set of particles we used is very small.
However it has been noted \cite{Bsh2} that, in the case of the compressibility factor at dimensions as 
high as $D=7$, the comparison of numerical results with 64 particles are in good agreement with simulations of systems of the order of thousands of particles. This may be confirmed in Fig. \ref{BComp}, which shows the comparison with our simulation data. Size effects are very small for $D=4$ and $D=5$. For $D=6$ these changes seems to be small but may be important. Therefore, we also made a simple examination of how the size of the sample affects the measurement of the value $g(X_{min})$ for the case $D=6$, for which we use the smaller samples. Considering not only the hypercubic primitive cell ($N_p=32$), but also the rectangular simulation cells where one of the sides of the simulation cell is composed by 2 primitive cells ($N_p=64$) and the case where two sides have 2 primitive cells ($N_p=128$), to estimate the changes for a density close to the freezing point ($\rho=1.3$). On the RDF one can notice first that, the larger sample is reflected in general as a smoother distribution. On the other hand, the evolution of the estimations of $g(X_{min})$ in time shows that the mean value converges with differences of the order of $2\%$ between systems of $N_p=32$ and $N_p=64$ particles, while the differences between $N_p=64$ and $N_p=128$ are less than $0.5\%$. It seems that as we increase dimensionality the growth of the system size affect less the equilibrium properties. For $D=7$ the comparison is possible only with simulation data in the fluid branch, however we believe the behavior of $g(X_{min})$ for larger simulation system will be close to the one we have obtained.

\begin{figure}
\includegraphics[scale=0.9]{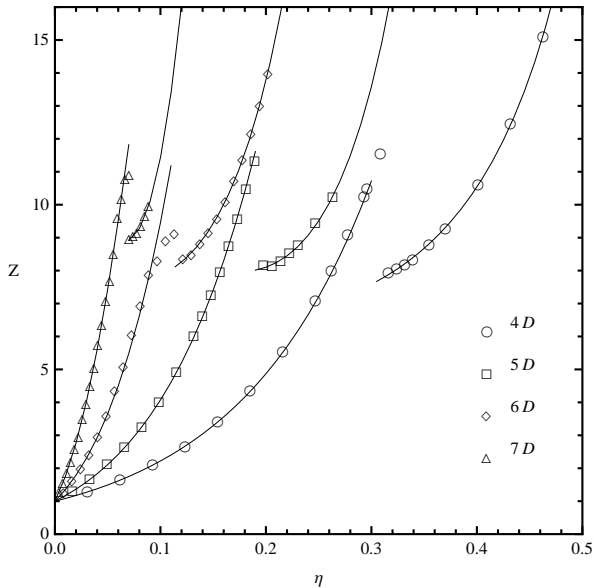}
\caption{Compressibility factor for $D$=4, 5, 6 and 7 as obtained by molecular dynamics simulations(dots) compared with reported semiempirical equations of state by Bishop et al. \cite{Bsh,Bsh2}(lines) and fits (Eq. (\ref{ZahllD}) and coefficients of Table \ref{TFC}) to the respective crystalline branches.}\label{ZS47D}
\end{figure}

For the whole range of densities, our values of the compressibility factor displayed in Fig. \ref{ZS47D},  
show for all examined dimensions a similar qualitative behavior to the $D=3$ case. 
Simulation data agree with reported equations of state up to $D=7$ for the fluid phase.
In particular for Fig. \ref{ZS47D} the continuous lines that reproduce the fluid phase, were chosen to be a Pad\'e[4,5] of the form:
\begin{equation}\label{Z45}
Z_{[4,5]}(\rho)= \frac{1 + p_1 \rho + p_2 \rho^2 + p_3 \rho^3 + p_4 \rho^4}{ 
 1 + q_1 \rho + q_2 \rho^2 + q_3 \rho^3 + q_4 \rho^4 + q_5 \rho^5}
\end{equation}
which is known to be accurate up to $D=9$ using the coefficients $\{p_i\}$ and $\{q_i\}$ determined by Bishop et al. \cite{Bsh,Bsh2}.

For the crystalline branch, and for dimensions $D>3$, we propose that one accurate fit is given by:
\begin{equation}\label{ZahllD}
 Z_{C}(\rho)=\frac{c_0+c_1\rho+c_2\rho^2}{1-\rho/\rho_{CP}}
\end{equation}
where $\rho_{CP}$ is the close packed density. To be compatible with reported equations of state, this proposal is expressed in terms of reduced density but may be rescaled to packing fraction. The fit of our simulation data gives the values of $c_0$, $c_1$ and $c_2$ shown in Table \ref{TFC}, and the values of $\rho_{CP}$ (or $\eta_{CP}$) were obtained from the lattice catalog of Nebe and Sloane \cite{NS}.
\begin{table}
\caption{Fitting parameters for the equation of state $Z_C$ for the crystalline branch.}\label{TFC}
\begin{ruledtabular}
\begin{tabular}{llllll} 
  &$\eta_{CP}$&$\rho_{CP}$&$c_0$&$c_2$&$c_2$\\ \hline 
4 & 0.61685 & 2.0 &14.6452&-18.6503&9.07211\\ 
5 & 0.46526 & 2.82843 &16.0717&-16.0419&6.36201\\ 
6 & 0.37295 & 4.6188 &11.7468&-7.02387&2.46827\\
7 & 0.29530 & 8.0 &17.3448&-9.86783&2.56618\\ 
\end{tabular}
\end{ruledtabular}
\end{table}

Other works use the functional form by Speedy \cite{RJS98}
\[
Z_{CSp}=\frac{D}{1-\rho/\rho_{CP}}-\frac{a_{sp}(\rho/\rho_{CP}-b_{sp})}{\rho/\rho_{CP}-c_{sp}}
\]
where the values of $a_{Sp}$, $b_{Sp}$, and $c_{Sp}$ are fitting constants. Both expressions are equally valid, and fit very well the crystalline branch. However, the proposed equation (\ref{ZahllD}) seems to yield to better agreement for the metastable crystal branch.
   
The structural properties are analog to the $D=3$ case, showing the same disorder-to-order transition once 
the density increases and the system changes from fluid to metastable solid phase. 

In the analysis of the values of $g(X_{min})$ for $D=4$ to $7$ shown in Fig. \ref{GXm47D}, it is remarkable that the analogy to a second order phase transition seems to be still valid for all 
examined dimensions. 
Therefore we applied the same numerical method used for $D=3$ in the estimation of the freezing points
for $D=4$ to $7$, i.e. through a numerical fit of two polynomials of order three to the mean values. 
The uncertainty was estimated taking into account the error bars of $g(X_{min})$.
The results of this procedure are in Table \ref{TF} under the label \textit{PF}. The estimated values
up to $D=5$ could  be compared with previous estimations by Michels and Trappenier \cite{MT}, Luban 
et al. \cite{LMM} and recent results by Van Meel et al. \cite{VMeel2}(see columns 1,2 and 3 in Table \ref{TF}). For all treated dimensions we compared also with the theoretical result of Xian-Zhi 
\cite{XZW}, which are based on a mean field cage theory, and may be summarized in the general expression:
\begin{equation}
 \rho_f=\frac{\rho_{CP} (1+2D)}{2+2D \rho_{CP} \sigma^D},
\end{equation}
Again the values of $\rho_{CP}$ were taken from reference \cite{NS}. The agreement with our results is good until $D=6$ and lie in the same order of magnitude up to $D=7$.

\section{Discussion.}
\label{sec4}

\begin{figure}
\includegraphics[scale=1]{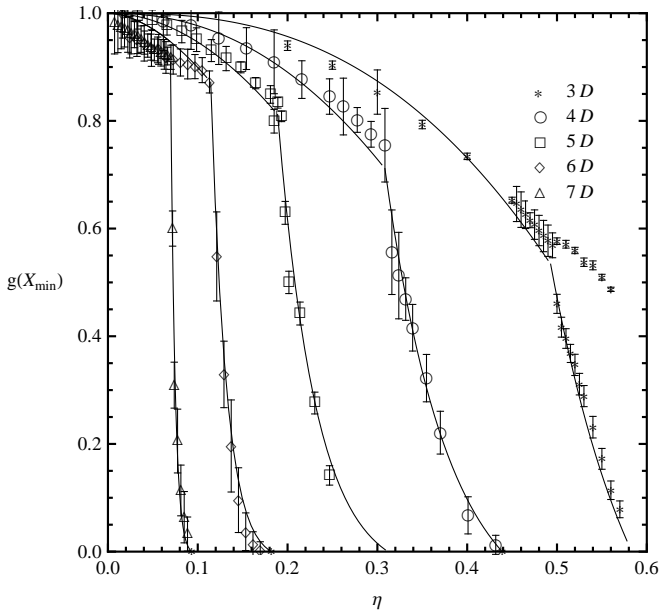}
\caption{Comparison of numerical simulation values of $g(Xmin)$ (dots) with proposed expressions (\ref{GMF}) and (\ref{GMC}) for $D$=3,\ldots ,7.}\label{GXm47D}
\end{figure}

Following the work of Rohrman and Santos \cite{RFA2}, it is possible to approximate the RDF of HHS fluids
in odd dimensions using a RFA method. Therefore by this method the changes on the first minimum of the RDF can be 
tracked as a function of the packing fraction and the dimensionality. Unfortunately, this procedure is
possible only numerically, but can give more information on what is observed in simulations. 
A numerical inspection  of the predictions of the RFA method for $D=5$ to $D=9$ suggests that the value of
$g(X_{min})$ may be expressed by a function of $\eta$ and $D$ which should comply with the following characteristics: 
$i)$ in the limit of low packing fractions the value of $g(X_{min})$ must tend to 1. $ii)$ The value of $g(X_{min})$ should decrease with density following a power law and $iii)$ the dimension modulates the width of the curve. 
A simple expression that meets these requirements is:
\begin{equation}\label{GMF}
 g(X_{min})=g_{MinL}(\eta,D)=1-D \eta^{\alpha},
\end{equation}
where the coefficient $\alpha$ may depend on $D$. 
Using this equation to fit the numerical predictions of the RFA method for $D$=3,5,7 and 9 
it is possible to establish that $\alpha$ must have a tendency that approximately follows an equation of the
form:
\[
%\alpha(D)= 1.484+2.922 e^{-0.28551 D}.
\alpha(D)= \alpha_0 D^{-\alpha_1},
\]
where the two parameters involved best fit the predictions with the values $\alpha_0=1.59243$ and $\alpha_1=0.5686$. The advantage of deriving these relations is that expression (\ref{GMF}) may be assumed 
to be valid also for even dimensions, where no predictions are available from the RFA method and
therefore may be used to compare all the simulation data obtained. Such comparison is shown in Fig. \ref{GXm47D},
where the Eq. (\ref{GMF}) was used to evaluate the continuous lines approaching the simulation points in
the fluid phase. Clearly, they provide us the good estimations both qualitatively and quantitatively for all of them.

The value of $g(X_{min})$ in the crystalline branch can not be evaluated yet from any theoretical scheme,
but examining with some detail the simulation data at even and odd dimension, one can note that the curves
present an exponential decay with density. It is also reasonable to expect that $g(X_{min})$ goes to zero at some finite packing fraction lower than the close packing $\eta_{CP}$. Therefore, we propose the following general expression:
\begin{equation}\label{GMC}
 g_{MinC}(\eta, D)=D^{-b(\eta+c)}+d
\end{equation}
where the parameters $b$, $c$ and $d$ depend on $D$ and may be obtained as the best fit to the simulation 
data, giving the values included in Table \ref{TCoef}.
\begin{table}
\caption{Fitting coefficients for $ g_{MinC}(\eta, D)$.}\label{TCoef}
\begin{ruledtabular}
\begin{tabular}{llll} 
d& b & c &d \\ \hline
3 & 7.559& -0.487& -0.425\\ 
4 & 10.422& -0.296& -0.131\\ 
5 & 20.715& -0.182& 0.048\\ 
6 & 30.999& -0.112& -0.049\\ 
7 & 42.107& -0.067&  -0.053\\ 
\end{tabular}
\end{ruledtabular}
\end{table} 

The changes observed for the parameters can be reproduced with exponential functions of $D$.
The best fitting expressions to the data in Table \ref{TCoef} are: 
\begin{eqnarray*} 
b(D)&=& 8.942 + 0.00461 e^{1.420 D}\\
c(D)&=& -2.059 e^{-0.485 D}\\
d(D)&=& -0.000945 - 7.082 e^{-D}\\
\end{eqnarray*}
It is remarkable to note that as $D$ increases, the parameter $b$ increases fast, indicating that the curve corresponding to the metastable crystalline states decreases faster for higher dimensions and at the limit $D\rightarrow\infty$ the curve changes in a discontinuous way indicating the colapse of the freezing and melting point to the same value.

Equations (\ref{GMF}) and (\ref{GMC}) can also  be used to estimate the freezing point, by solving
the equation  $g_{MinL}(\eta,D)=g_{MinC}(\eta,D)$. Clearly the solution can not be 
algebraic but anyway provides a semiempirical method to determine the freezing point for arbitrary dimensions. 
A list of estimated values for the freezing point computed whit the simple the polynomial fit used before (\textit{PF}) and  the universal relations (\textit{UR}) given by Eqs. (\ref{GMF}) and (\ref{GMC}) are presented in Table \ref{TF}.\\

\begin{table}
\caption{Comparison of freezing $(\eta_F)$: Previous reported values, theoretical prediction from Mean Field Cage Theories (MFCT), recent estimations by Van Meel and computed values from Polynomial Fitting (PF) and derived Universal Relationship (UR).}\label{TF}
\begin{ruledtabular}
\begin{tabular}{lccccc} 
\multicolumn{1}{c}{$D$}&\multicolumn{5}{c}{$\eta_F$}\\ 
   &Prev. Sim.     & MFCT\cite{XZW} & Van Meel \cite{VMeel} & PF     & UR     \\  \hline
3  &0.494\cite{AW3}& 0.494          & 0.494  & 0.488(5)  & 0.492\\ 
4  &0.308\cite{MT} & 0.308          & 0.288  & 0.304(1)  & 0.308\\  %
5  &0.194\cite{MT} & 0.169          & 0.174  & 0.190(1)  & 0.189\\ 
6  &               & 0.084          & 0.105  & 0.114(2)  & 0.114\\ 
7  &               & 0.039          &        & 0.0702(2) & 0.0696\\ 
8  &               & 0.017          &        &           & 0.0427\\ 
\end{tabular}
\end{ruledtabular}
\end{table}

\begin{table}
\caption{Coexistence reduced pressure $(p^*_F)$ computed from the CS equation of state for $D=3$ and Pad\'es form Bishop et al. for $D=4,5,6,7$}\label{TP}
\begin{ruledtabular}
\begin{tabular}{llll} 
\multicolumn{1}{c}{$D$}&\multicolumn{3}{c}{$p^*_{F}$}\\ 
  &Prev. Sim. & PF      & UR\\ \hline \hline
3 &11.779     & 11.202 & 11.668\\ 
4 &11.418    &  11.008 & 11.469\\ 
5 &14.315    &  13.433 & 13.184\\ 
6 &           & 16.668 &17.0318\\ 
7 &           & 22.597 & 22.1569\\ 
\end{tabular}
\end{ruledtabular}
\end{table}

\begin{table}
\caption{Melting $(\eta_M)$ packing fraction (By extrapolation of $p^*$ to crystalline branch)}\label{TM}
\begin{ruledtabular}
\begin{tabular}{lllll} 
\multicolumn{1}{c}{$D$}&\multicolumn{4}{c}{$\eta_M$}\\ 
  & Sim.   & Van Meel \cite{VMeel} & PF      & UR\\ \hline 
3 & 0.545  & 0.545 & 0.537 & 0.542\\ 
4 & 0.337  & 0.337& 0.368 & 0.374\\ 
5 & 0.206  & 0.206 & 0.242 & 0.240\\ 
6 & 0.138  & 0.138 & 0.146 & 0.147\\
7 &        & & 0.086 & 0.085\\ 
\end{tabular}
\end{ruledtabular}
\end{table}

The results summarized in Table \ref{TF} show that the use of semiempirical equations improve the estimation
of the freezing point, at least for $D=4$ and $D=5$ where previous simulations and theoretical approaches are
available. For $D>6$ our estimation always lies above the theoretical predictions but no other
simulations results are available. \\

In addition, with the estimated freezing points we can extrapolate the melting point and transition pressure
using the equations of state given in Eqs. (\ref{Z45}) and (\ref{ZahllD}) for fluid and crystal phases, and
using the condition of equal pressure in the coexistence region. The resulting values are in Tables \ref{TP} and \ref{TM}, where, although being extrapolations, may give very good results compared with previous
simulations up to $D=6$. Again an improvement in the approach may be remarked from the semiempirical
solution. This means that the scheme of chosen equations of state and the implicit equation for the 
freezing point is consistent with present available data.
 
It is important to remark that the freezing packing fractions, estimated using any of the strategies 
described, change as a function of $D$ following an exponential decay, and fit very precisely the function
$\eta_F(D)=2.103 e^{-0.482 D}$. It is not clear whether it is possible to ensure that this relation will 
be the same for $D$ out of the examined range. In particular for $D=1$ and $D=2$ it extrapolates to $\eta_F(1)=1.298$ and 
$\eta_F(2)=0.802$, which do not correspond to the known values ($\eta_F=1$ for $D=1$ and $\eta_F=0.69$ for $D=2$). On the other hand, the intersection of relations (\ref{GMF}) and (\ref{GMC}) gives the values $\eta_F=1.215$ and $\eta_F=0.756$ respectively.
Also for $D>7$ the tendency can not be settled as the definitive one.

At this point, we have presented evidence that the use of the properties of the
radial distribution function at the first minimum can be regarded as an empirical method to estimate the freezing
point in HHS fluids.
However, it is not clear why the position and the height of the first minimum of the
RDF in the crystal and 
the fluid branches could be connected with the freezing point for all
dimensionalities. The answer is not at all 
simple, but some properties of the three dimensional fluid may give some insights. 

In a hard sphere system the fluid-solid transition happens, either when the rise or
decrease of the free volume allows 
the formation or destruction of cages. The properties of such cage formation
determine whether the fluid form crystals or glasses. In fact, the concept of
dynamical formation of cages has been already used to approach the phase diagrams of
hard disks and spheres \cite{ASK08}. 
In our case our MD simulations began from spheres placed in an FCC lattice (for the
crystal branch) and high packed random states (metastable branch), and we control
the packing fraction by changing the diameter of the spheres. As is usual, 
this kind of simulations do not allow to achieve the coexistence branch because of
the finite size of the samples \cite{Skg}, therefore the crystal
branch obtained includes a metastable region between the melting and the freezing
points where cages must be present
preserving some of the crystalline order observed in the RDF.
When the density decreases 
the state where those crystalline structures melt could correspond to a state
where particles leave cages, and this happens very close to the freezing density.

The value of $X_{min}$ has been usually related with the mean number of first
neighbors. The integral of the RDF in a 
spherical volume of radius $X_{min}$ times the density should, in principle,
approach this number. In fact, the
more packaged the system is, the better approach we get (at least for the crystal
branch). In this
context $X_{min}$ may be interpreted as the mean radius of a spherical volume that
contains the first neighbors around a given sphere. 

What we observe in Fig. \ref{Xm3D} is that the states in the crystalline
metastable branch approaching the freezing point are still FCC structures where first
neighbors are confined in average in a sphere of radius lower than $\sqrt{2}$ times the
diameter, a volume in which an ideal close-packed crystal may contain up to the second neighbors.
It seems that once this volume surpasses this value the metastable states quickly ``melt''. Interestingly
enough the packing fraction at which this ``melting'' occurs  coincides with that of the freezing point.

Recently Kumar and Kumaran \cite{VSK05} found, using a Voronoi neighbor
statistics, the change on the first 
neighbor number $C_1$ as a function of the packing fraction for the hard-sphere
system. They generated hard sphere structures using the NVE Monte Carlo (MC) and
annealed Monte Carlo (AMC) methods, to obtain configurations
in two branches, one where crystallization is allowed and another where not. A plot of
$C_1$ as a function of $\eta$, given in
the last reference, shows a behavior similar to our Fig. \ref{Xm3D}, where for the
fluid phase $C_1$ has a smooth decay when the packing fraction increases, starting
from values above $15.5$ at low densities. Once the number of neighbors is around
$14.5$
the structures obtained by MC begin the crystallization and show a sudden decrease
indicating the freezing transition. They obtain a
value of $C_1$ for crystalline configurations close to $14$ instead of the expected $12$
because a slight perturbation of the position in the centers of a pair of molecules
may transform a vertex in a surface introducing second neighbors to the Voronoi
count. Annealed structures do not crystallize and lead to random dense structures
that may be compared with our metastable fluid branch.

\begin{figure}
\includegraphics[scale=0.5]{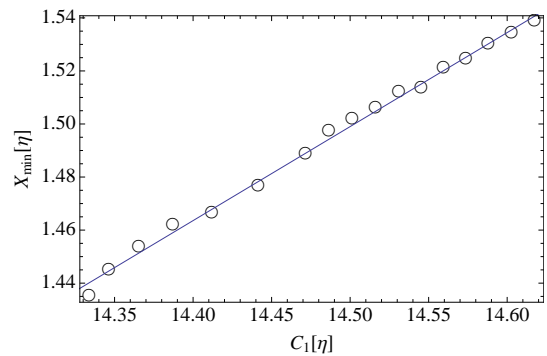}
\caption{Comparison between values of the first neighbor number obtained in
\cite{VSK05} and the radius of first minimum for the liquid-liquid metastable
branch.}\label{Kum-1}
\end{figure}

\begin{figure}
\includegraphics[scale=0.5]{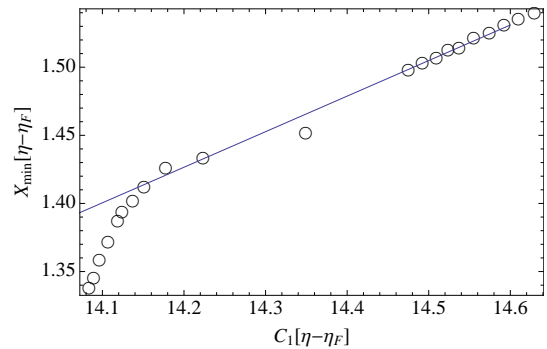}
\caption{Comparison between values of the first neighbor number obtained in
\cite{VSK05} and the radius of first minimum for the liquid-crystal
branch.}\label{Kum-2}
\end{figure}

To inspect the relation among $C_1$ in reference \cite{VSK05} and $X_{min}$ we
produced Fig. \ref{Kum-1} and Fig. \ref{Kum-2}, 
relating both variables. In Fig. \ref{Kum-1} we put together both values
assuming that the AMC simulations generate configurations in the same metastable
fluid branch our work does. From the figure a proportional relation between
both quantities in the region of packing fractions close to the freezing point is clear. For
configurations obtained from NVE (MC) one may obtain Fig. \ref{Kum-2} comparing
with the crystal branch of our $X_{min}$ results, where we note that for comparison we 
moved each function to center both in their respective freezing packing fraction,
which are close, but are different estimations. Despite the differences on simulation 
methods it is possible to presume that in both cases (more clearly in the metastable
branch) there is a linear proportion between $C_1$ and $X_{min}$ in the vicinity of the
freezing point. The sudden change on $X_{min}$ is proportional to the change
on the number of first neighbors, indicating that any remaining cages in the
metastable states are opening.

If we put together the last elements we can interpret that the changes in $X_{min}$ and
$g(X_{min})$ close to the  freezing transition are directly related with the critical 
changes on the cages present in the system. A critical number of first neighbors 
and a critical radius define either the formation or destruction of the cages. In 
other words the dynamical formation of cages, in the context cited by Kreamer and 
Naumis \cite{ASK08}, may give the effect observed in the first minimum of 
the RDF.

For higher dimensions something similar may be happening but without a doubt in a
more complicated context. For example,
while the $D_4$ and $D_5$ lattices seems to be the densest packings for $D=4$ and
$D=5$, in dimension $D=6$ the $E_6$ lattice is more compact than $D_6$
\cite{JHC98}. In 2006 Skoge et al. \cite{Skg}, found that for
random 
jammed states the first minimum of the RDF for $D=3,4,5$ is close to where the
cumulative coordination number equals the kissing number  of the densest lattice,
giving some clues that packing phenomena may present analog properties at least up 
to $D=5$. However, in a more recent work van Meel et al. \cite{VMeel}
studied the effect of frustration on random packed structures in $D=4$ finding that
in this system crystallization is slower than in $D=3$ because the geometrical
frustration is surprisingly stronger than in $D=3$ and that the similarity between
the kissing number and the number of first
neighbors can be explained by a wide first peak of the RDF, that accommodates
nonkissing neighbors in polytetrahedral clusters. 
Beyond these achievements the problem of dynamical formation of cages in the freezing
and melting transition is not yet examined
deeply and neither the relation  with the first minimum of the RDF. In our
point of view it is an still open problem that deserves some future work.

\begin{figure}
\includegraphics[scale=1]{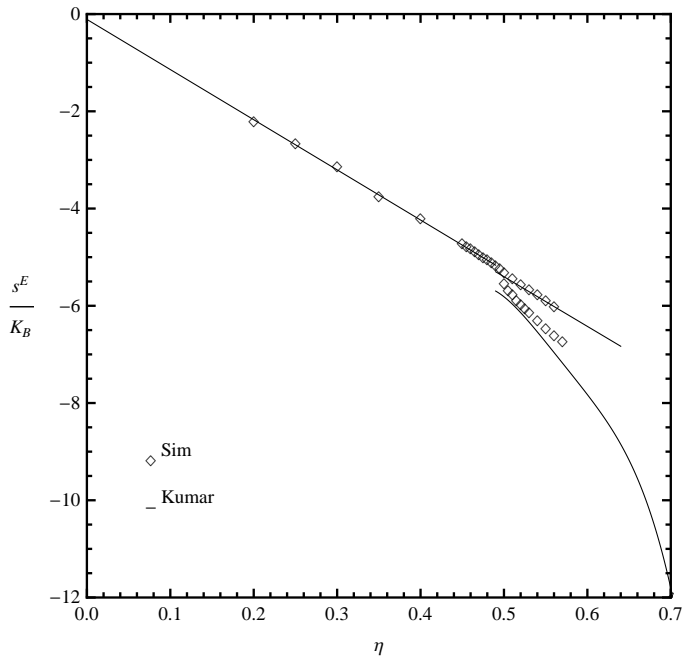}
\caption{Comparison of hard-spheres excess entropy as computed from Eq. (\ref{entropy}) using values of $X_{min}$ obtained from simulations (labeled Sim), results from Voronoi cell distribution from Ref. \onlinecite{VSK05b} (labeled Kumar).}\label{ESHS}
\end{figure}

In addition it is also possible to trace a path to a thermodynamic
interpretation of the observed effects. This may be achieved through the examination of
the entropy. In another recent work of Kumar and Kumaran \cite{VSK05b}, the configurational entropy 
of hard spheres was examined using the volume distribution of Voronoi cells. 
There they found a relation between the information entropy of the distribution and the
excess entropy of the hard-sphere fluid, which we have approximately fitted in Fig. \ref{ESHS}.
One may note that such results can be very well reproduced from our results if we define an excess free-volume-like entropy per
particle as:
\begin{equation}
 \frac{s^E}{K_B}=\lambda \log(V_{min}(1-\eta)/V_0),
\label{entropy}
\end{equation}
where $V_{min}=(4 \pi/3) X_{min}^3$ is the spherical volume corresponding to $X_{min}$ and therefore $V_{min}(1-\eta)$ is the mean available volume of a sphere within the first neighbor cell. The parameter $\lambda$ is chosen to be $\lambda=3$ in accordance with the proportionality constant between the information an excess entropies found by Kumar and Kumaran and $V_0$ left as free parameter with a value $V_0=40.45$. From the comparison, shown also in Fig. \ref{ESHS},
it is clear that both results must be related and therefore this definition is fully compatible with the
information entropy derived from Voronoi statistics. 
% The parameter $\lambda=3$ is also used in
% the reference to relate the information entropy and the excess entropy derived from analytical approximations to the equation o state, the whole comparison is also in Fig. \ref{ESHS}.
From here we can derive two conclusions for the system $D=3$: first, Fig. \ref{Xm3D} and Fig. \ref{GXm3D} 
may be taken as two expressions of the freezing transition in the entropy-density plane; second, as a consequence, the changes at
freezing density are not continuous because an entropy gap appears between the last metastable crystalline state and the first fluid state. Kumar and Kumaran \cite{VSK05b} report for the entropy gap an estimation of $\Delta s/K_B\sim0.92$ which corresponds, using equation (\ref{entropy}), in the $X_{min}$ space to a gap $\Delta X_{min} \sim 0.15$. The gap we observe in our simulations is lower than $\Delta X_{min} \sim 0.05$. Both of them are small in the scale of $X_{min}$ and of the order of the error bars, allowing a good accuracy for the estimation of the freezing density if a continuous change is assumed at the freezing point. Since Kumar and Kumaran use the same form of Eq. \ref{entropy} for $D=2$ and $D=3$ with $\lambda=D$,  we believe the same situation will apply for $D>3$ and  Eq. (\ref{entropy}) could be generalized for arbitrary $D$  (with $V_0$  a function of $D$) as
\begin{equation}
 \frac{s^E(D)}{K_B}=D \log(V_{min}(1-\eta)/V_0(D)).
\label{entropyD}
\end{equation}
This could be the reason why the estimation method for the freezing point through $X_{min}$ and $g(X_{min})$ seems to  work also for the examined dimensions. We plan to perform a further and deeper
analysis of the entropy properties of hypersphere systems in future work.

\section{Summary and closing remarks.}
\label{sec5}

We have studied the problem of freezing in HHS fluids through molecular dynamics 
simulation from $D=3$ to $D=7$, obtaining data for the compressibility factor and the RDF 
in dense liquid states close to freezing and crystal metastable states, below the melting density.
We found that the changes with density of the first minimum of the RDF, in all examined dimensions,
follow a tendency  
%%%%Antes%%%%
%that resembles a second order phase transition
%%%%Ahora
where the three branches seems to converge to a single point.
The packing fraction where this transition takes place
is a good approximation to the previously estimated freezing points at  $D=3$, $4$ and $5$, giving
more elements to affirm such values are close to the correct ones. Assuming that the same structural changes 
occur in higher dimensions we applied the method in our results for $D=6$ and $D=7$, obtaining
new estimations of the freezing packing fractions that could be in the future confirmed with different 
theoretical and numerical techniques. The proposed method was also formulated in a semiempirical scheme,
using the RFA method of Rohrman and Santos \cite{RFA2} to obtain the first minimum of the RDF numerically
in the fluid phase and finding the intersection with empirical relations in the crystal phase.
General relations for the value of the RDF in the first minimum, as a function of the dimension and the packing fraction were proposed, giving also very good approximations to known values for the freezing points.

Although the sets of particles used in this work are small, we are confident that the results describe the overall system behavior. Previously it was noted that for dimension $D=7$, the equation of state is well described even for a small set of particles ($N=64$). A simple comparison of the $g(r)$ properties shows that the influence of the system size is reflected as a slower convergence and a more noisy numerical result, but the mean value of $g(X_{min})$ seems to be consistent for different sizes of the system. However, a more detailed analysis may clarify this conjecture.   

An entropy analysis reveled that the properties of $X_{min}$ and $g(X_{min})$ may be related
with the changes in the excess entropy of the system and that a small gap
between the fluid and the crystalline metastable branches exists, but the size of it is
small enough to allow the use of the proposed numerical method to obtain the estimations of the freezing density with good accuracy.

Finally, it is worth to mention that the method of tracking the minimum of the RDF is reliable for the examined cases, but as we increase the dimensionality there is an effect of flatness in the fluid RDF and sharpness in the the crystal RDF, collapsing all minima close to $1$  in the fluid phase and close to $0$ in the crystal phase, and therefore, the accurate determination of the position and height of either the maximum or minimum becomes more difficult.

\acknowledgments
We acknowledge the financial support of DGAPA-UNAM through the project IN-109408-2.  One of us, C. D. E. acknowledge to CONACyT for financial support through the grant 171940. We also want to thank Prof. Mariano L\'opez de Haro for helpful comments.

%\bibliography{Articulo3.7.6.bib}
%\end{document}

%Merlin.mbs v4.21 2009-07-09.
%

\end{document}